\documentstyle[aps,twocolumn,tighten,epsfig]{revtex}
\begin{document}
\draft
\title{SYMMETRY BREAKING WITH A SLANT: \\
       TOPOLOGICAL DEFECTS AFTER AN INHOMOGENEOUS QUENCH}

\author{Jacek Dziarmaga$^{(1)}$,
        Pablo Laguna$^{(2)}$ and
        Wojciech H. Zurek$^{(3)}$}
\address{ $^{(1)}$ Institute of Physics, Jagiellonian University,
                   Reymonta 4, 30-059 Krak\'ow, Poland\\
          $^{(2)}$ Dept. Astronomy \& Astrophysics and CGPG,
                   Penn State University, University Park,
                   PA 16802, USA\\
          $^{(3)}$ Theoretical Astrophysics T-6, MS-B288,
                   Los Alamos National Laboratory, Los Alamos,
                   NM 87545, USA}

\maketitle
\tighten

\begin{abstract}
\noindent We show that in second-order phase transformations induced by
an inhomogeneous quench the density of topological defects is drastically
suppressed as the velocity with which the quench propagates falls below
a threshold velocity. This threshould is approximately given by the ratio
of the healing length to relaxation time at freeze-out, that is at the
instant when the critical slowing down results in a transition from the
adiabatic to the impulse behavior of the order parameter.
\end{abstract}


In a uniform cosmological phase transition characterized by an order
parameter with a non-trivial homotopy group, creation of topological defects
is virtually inevitable. As pointed out by Kibble \cite{kibble} some years
ago, different domains shall select different low-temperature vacua, leading
to irreconcilable differences which condense out into topological defects.
Thus, special relativity implies disorder in the context of Big Bang,
yielding a useful lower bound on the initial density of defects. An analogous
situation is also encountered in the condensed matter setting \cite{zurek1},
suggesting the possibility of experimental investigation of a cosmological
scenario. For this case, however, a lower bound based on the light-cone
causal independence is no longer useful. In condensed matter systems (and
most likely also in the cosmological second-order phase transitions), the
initial density of defects has to be computed through arguments which rely
on the dynamics of the order parameter \cite{zurek1,zurek2}.

The estimate of defect density proposed by one of us relies on the
observation that, as a result of critical slowing down, the state of the
system which crosses the critical region at a finite pace will inevitably
cease to keep up with the changes of thermodynamic parameters at some point
sufficiently near the critical temperature \cite{zurek1,zurek2}. In a
homogeneous quench, this will happen everywhere at the instant when the
characteristic time $\epsilon/\dot \epsilon$, at which changes in the
mass term of the effective Landau-Ginzburg (LG) potential
$ V(\phi) = (\phi^4 - 2 
\epsilon(t) \phi^2)/8 $
become faster than the relaxation time, i.e. $\tau =\tau_0|\epsilon|^{-1}$
in the overdamped case. When $\epsilon(t) \simeq t/\tau_Q$, this leads to a
freeze-out time $\hat t \simeq (\tau_Q \tau_0)^{1/2}$. At $t=-\hat{t}$
the order parameter $\phi$ becomes too sluggish to keep up with the evolving
shape of the effective potential. An evolution dominated by dynamics driven
by $V(\phi)$ shall resume $\hat t$ after the phase transition, which takes
place when $\epsilon(t=0) = 0$. The interval $[-\hat t, + \hat t]$ represents
the impulse stage, namely the
period during which the effect of $\partial V\partial\phi$ is negligible,
although fluctuations and damping continue to matter. At $ \hat t$, 
$\hat \epsilon \equiv \epsilon (\hat t) = (\tau_0 /  \tau_Q)^{1/2}$, which leads
to the estimate of the corresponding healing (or correlation) length
$\hat \xi= \xi_0 ~ (\tau_Q / \tau_0)^{1/4} $. This freeze-out healing
length constrains the domain size
relevant for defect formation. The corresponding relaxation time is
$\hat \tau = (\tau_Q \tau_0)^{1/2}$, and the characteristic phase ordering
speed is $\hat v = \hat \xi/\hat \tau = v_0 ~ (\tau_0 /\tau_Q)^{1/4}$, where
$ v_0 = \xi_0/\tau_0$. This overdamped LG example
can be generalized \cite{zurek2}.

Homogeneous quenches are a convenient idealization and may be a good
approximation in some cases. However, in reality, the change of thermodynamic
parameters is unlikely to be ideally uniform. Thus, the mass parameter
$\epsilon(t,\vec{r})$, varying in both time and space, must be considered in
defect formation. As a consequence, locations entering the broken symmetry
phase first could then communicate their choice of the new vacuum as the
phase ordered region spreads in the wake of the phase transition front. When
inhomogeneity dominates, symmetry breaking in various, even distant,
locations is no longer causally independent. The domain where the phase
transition occurred first may impose its choice on the rest of the volume,
thus suppressing or even halting production of topological defects.

Experiments carried out in $^3He$ \cite{Helsinki,Grenoble}, where a small
volume of superfluid is re-heated to normal state, and subsequently rapidly
cools to the temperature of the surroundings, are a good example of an
inhomogeneous quench: The normal region shrinks from the outside. Yet,
topological defects are created, thus suggesting that the phases of distinct
domains within the re-heated region are selected independently. Even in $^4He$,
where the transition can be induced by a change of pressure, it is difficult
to rule out the possibility that a quench may be somewhat non-uniform, thus
causing decrease of the density of defects, which could explain the recent
evidence of non-appearance of vortices \cite{Dodd}.

Here we consider two idealized cases of inhomogeneous quenches: (i) a 
{\it shock wave} in which $\epsilon(t,\vec{r})$ is a propagating step in space
and (ii) a {\it linear front} in which $\epsilon(t,\vec{r})$ linearly
interpolates between the pre- and post-transition values. For both scenarios,
we investigate the threshold velocity $v_t$ at which phase ordered region
expands behind the $\epsilon=0$ critical point. This velocity is given by
the ratio of the healing length to relaxation time set by the dynamics. In an
inhomogeneous quench, $v_t$ will be --- as it was anticipated early on
\cite{zurek1} --- of the order of $ \hat v$.

In an insightful paper directly stimulated by the $^3He$ experiments, Kibble
and Volovik have argued that the initial density of defects should conform
with the homogeneous quench estimates of \cite{zurek1,zurek2} when the
velocity $v$ of the inhomogeneous quench front exceeds $\hat v$; on the other
hand, the initial density of defects should be suppressed as $v/ \hat v$ for
the case of a slowly spreading phase transition \cite{kv}. We find that
$\hat v$ indeed plays a crucial role. However, our study shows that the
suppression is even more dramatic. When $\hat v > v$, essentially no defects
appear.

\paragraph*{\bf  SHOCK WAVE. }

To begin with, we consider a decay of a false symmetric vacuum to a
true symmetry broken phase in a one-dimensional dissipative $\phi^4$
model. In dimensionless units,

\begin{equation}\label{varphi4}
\ddot{\phi} +
\eta\,\dot{\phi} \; -
\; \phi'' \; +
\frac{1}{2} \;(\phi^3-\epsilon_0\phi) = 0\;,
\end{equation}
where $\phi(t,x)$ is a real order parameter and $\epsilon_0$
is a positive constant. We look for a solution $\phi(t,x)$ which
interpolates between $\phi(t,-\infty)=-\sqrt{\epsilon_0}$ and
$\phi(t,+\infty)=0$. Such a solution can not be static; it is a
stationary half-kink

\begin{equation}\label{half-kink}
\phi(t,x)=-\sqrt{\epsilon_0}
\left(1+\exp{\left[\frac{\sqrt{\epsilon_0}(x - v_t^s t)}
                        {2(1-(v_t^s)^2)^{1/2}}\right]}\right)^{-1}
\end{equation}
moving with characteristic velocity

\begin{equation}\label{V}
v_t^s=\left[1+\left(\frac{2\eta}{3\sqrt{\epsilon_0}}\right)^2\right]^{-1/2}
\stackrel{\eta\rightarrow\infty}{\approx}
\frac{3\sqrt{\epsilon_0}}{2\eta}                               \;\;.
\end{equation}

Our shock wave quench is a sharp pressure front propagating with velocity $v$.
That is,

\begin{equation}\label{model}
\ddot{\phi} +
\eta\,\dot{\phi} \; -
\; \phi'' \; +
\frac{1}{2} \;(\phi^3-\epsilon(t,x)\,\phi) = 
\vartheta(t,x) \;\;,
\end{equation}
where

\begin{equation}\label{a}
\epsilon(t,x) \; = \epsilon_0 \; \mbox{Sign}(t-x/v)
\end{equation}
is the space-time dependent effective mass and $\vartheta(t,x)$ is a Gaussian
white noise of temperature $\theta$ with correlations

\begin{equation}\label{noise}
\langle \vartheta(t,x)\;\vartheta(t',x') \rangle \; =
\;2 \; \eta \; \theta \; \delta(t-t') \; \delta(x-x')  \;\;.
\end{equation}
There is a unique vacuum ($\phi=0$) to the right of the
propagating front ($x>vt$), and the symmetry is broken
($\phi=\pm \sqrt{\epsilon_0}$) behind the front ($x<vt$). 

The field in the $\phi=0$ vacuum does not respond instantaneously to the 
passing front. There are two qualitatively different regimes:

   1) $v>v_t^s$, the phase front propagates faster than the false vacuum
can decay. The half-kink (\ref{half-kink}) lags behind the front (\ref{a});
a supercooled symmetric phase grows with velocity $v-v_t^s$. The supercooled
phase can not last for long; it is unstable, and the noise $\vartheta$ makes
it decay into one of the true vacua. Since the noise does not have any bias,
the decay results in production of kinks.

   2) $v<v_t^s$, the phase front is slow enough for a half-kink to move in
step with the front, $\phi(t,x)=H_{v}(x-vt)$. The symmetric vacuum decays
into one definite non-symmetric vacuum, the choice is determined by the
boundary condition at $x\rightarrow -\infty$. No topological defects are
produced in this regime. To make sure that the field can not flip, we
must check if the stationary solution $H_v(x-vt)$ is stable against small
perturbations.

    We investigate the stability in the $\eta\rightarrow\infty$ limit
when the system is overdamped and the $\ddot{\phi}$ term in
Eq.(\ref{model}) can be neglected. $H_v$ is most likely to be unstable
for $v=v_t^{s-}$. We use the ansatz

\begin{equation}\label{ansatz}
\phi(t,x)=H_{v_t^s}(x-v_t^s t)+
          f(x-v_t^s t)\exp\left(\Gamma t-\frac{\eta\, v_t^s x}{2}\right) \,.
\end{equation}
The eigenequation turns out to be time-independent;

\begin{eqnarray}\label{eeq}
&-&\Gamma\;\eta \;f(x) =       \nonumber\\
&-&f''(x) +\left[  \frac{\epsilon_0}{2}\mbox{Sign}(x) +
                   \frac{\eta^2 (v_t^s)^2}{4} +
                   \frac{3}{2} H_{v_t^s}^2(x)\right]\; f(x) \;\;.
\end{eqnarray}
The ``potential" in the square brackets is positive definite for
$v_t^s=3\sqrt{\epsilon_0}/2\eta$. This proves the stability of $H_{v_t^s}$
at sufficiently low noise temperature $\theta$.

   In the opposite $\eta\rightarrow 0$ limit the half-kink (\ref{half-kink})
just below the threshould $v_t^s\approx 1$ becomes a step function
$H_{v_t^s}(x-v_t^st)\approx\sqrt{\epsilon_0}[-1+\mbox{Sign}(x-v_t^st)]/2$.
The potential on the r.h.s. of Eq.(\ref{eeq}) is again positive for
any $x$.

   In summary, no topological defects are produced for $v<v_t^s$. At $v=v_t^s$,
there is a sudden jump in the density of defects left behind the shock wave.
As $v$ increases above $v_t^s$, the density saturates at a value characteristic
for an instantaneous uniform quench with
$\epsilon(t,x)=\epsilon_0\;\mbox{Sign}(t)$. With increasing $\theta$,
the discontinuity at $v=v_t^s$ will be softened. For $v\gg v_t^s$, where the
quench is effectively homogeneous, the density of defects will grow
logarithmically with $\theta$ \cite{lythe}.

\begin{figure}[h]
\leavevmode
\epsfxsize=3.2truein\epsfbox{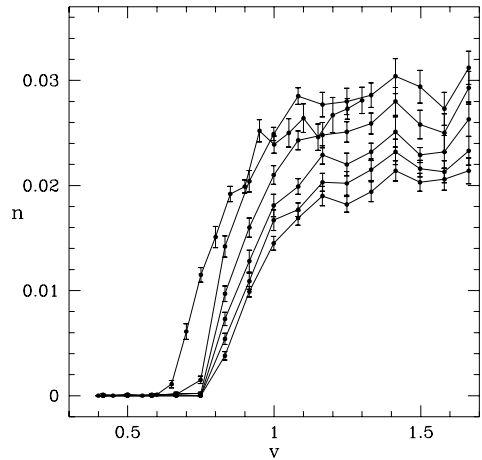}
\caption[figure1]{\label{fig1}
Density of kinks $n$ as a function of velocity $v$ for the shock wave
(\ref{a}) with $\eta=\epsilon_0=1$ (overdamped system).
In this overdamped regime, the predicted threshold velocity is $v_t^s=0.83$.
The plots from top to bottom correspond to $\theta=10^{-1},\, 10^{-2},\,
 10^{-4},\, 10^{-6},\, 10^{-8},\,10^{-10}$. At low $\theta$, we get a clear 
cut-off velocity at $v\approx 0.8$, which is consistent with the prediction. }
\end{figure}

These expectations are borne out by the numerical study of kink formation
which uses the same code as in Ref.~\cite{laguna}. We illustrate them
in Fig.~\ref{fig1} for $ \eta = \epsilon_0 = 1 $. For all but the
highest temperature $\theta = 0.1$, there are essentially no kinks produced
in quenches with the velocity of less than 0.8, which is in good agreement
with our analytic estimate $v_t^s = 0.83$. However, for the highest temperature,
defects appear at a sub-threshold velocity. We note that at this temperature
potential barrier separating the two minima of the LG potential is comparable
with $\theta$.


\paragraph*{\bf  LINEAR FRONT. }

Let us consider now the linear inhomogeneous quench

\begin{eqnarray}\label{model2}
&&\ddot{\phi}+\eta\dot{\phi}-\phi''+
  \frac{1}{2}(\phi^3 -\epsilon(t,x)\,\phi)=\vartheta(t,x) \;\;,
  \nonumber\\
&&\epsilon(t,x) =  \left\{
  \begin{array}{lrcr}
  -\epsilon_0           \quad &   \epsilon_0v\tau_Q \le &  x-vt      &                        \\
   \frac{vt-x}{v\tau_Q} \quad &  -\epsilon_0v\tau_Q \le &  x-vt \le  &  \epsilon_0v\tau_Q  \\
   \epsilon_0           \quad &                         &  x-vt \le  & -\epsilon_0v\tau_Q
\end{array}
\right.
\end{eqnarray}
We assume that the linear part of $\epsilon(t,x)$, namely the interval
$-\epsilon_0v\tau_Q \le x-vt \le \epsilon_0v\tau_Q$, is much wider
than the healing length to the left of the front,
$2\epsilon_0v\tau_Q \gg 1/\sqrt{\epsilon_0}$. If not then the shock
limit (\ref{a}) applies.

   In the absence of noise, the propagating linear front is followed by a
stationary half-kink $\phi(t,x)=h_{v}(x-vt)$. This half-kink lags a
distance $\delta x$ behind the front. $\delta x$ can be estimated by similar
arguments as those which led to Eq.(\ref{V}). At $\delta x$ behind
the front $x=vt$ the mass parameter is
$\epsilon_{\delta x}=\delta x/v\tau_Q$.
The replacement $\epsilon_0\rightarrow\epsilon_{\delta x}$ in
Eq.(\ref{V}) gives a velocity $v_t(\delta x)$ the half-kink would propagate
with if it were at $\delta x$. The velocity increases with $\delta x$.
The half-kink gets stuck at such a $\delta x$ that this velocity is equal
to the actual front velocity $v$, $v_t(\delta x)=v$. This takes place at
$\delta x=16\eta^4v^5\tau_Q/81(1-v^2)^2$, which grows quickly with $v$.

  When $v$ is greater than $v_t^s$ in Eq.(\ref{V}),
$\delta x>\epsilon_0v\tau_Q$ and the half-kink does not stay in the
linear regime. It enters the $\epsilon=\epsilon_0$ area where it moves
forward with velocity $v_t^s<v$. Like in the shock limit the supercooled
phase grows at a constant rate and decays giving rise to kinks.

  When $v<v_t^s$, the half-kink remains confined in the linear regime.
Even in this case, for $v$ greater than certain threshould $v_t$, the width
$\delta x$ of the supercooled phase may be large enough for this phase to be
unstable. A half-kink $h_v(x-vt)$ confined to the linear regime satisfies

\begin{equation}\label{UV}
(1-v^2)\;h''_{v}(x)+\eta\;v\;h'_{v}(x)-
\frac{x}{2v\tau_Q}\;h_{v}(x)-\frac{1}{2}h^3_{v}(x)=0 \;\;.
\end{equation}
We rescale $x=c_1\;\tilde{x}$ and $h_v=c_2\;\tilde{h}$ in Eq.(\ref{UV}),
then set $c_1=[2\;v\;\tau_Q\;(1-v^2)]^{1/3}$ and $c_2=\sqrt{c_1/v\tau_Q}$
to obtain the rescaled equation

\begin{equation}\label{UVbis}
L_1[\tilde{h}](\tilde{x}) \equiv
\tilde{h}''+p\;\tilde{h}'-
\tilde{x}\;\tilde{h}-\tilde{h}^3=0 \;\;,
\end{equation}
with primes now denoting derivatives w.r.t. $\tilde{x}$. Eq.~(\ref{UVbis})
has a single parameter
\begin{equation}\label{p}
p=\frac{ 2^{1/3} \eta \,v^{4/3} \tau_Q^{1/3} }{ (1-v^2)^{2/3} } \;\;.
\end{equation}
The half-kink $h$ becomes unstable at a threshold $p=p_t$. At this critical
$p_t$, $\tilde{h}$ has a zero mode $\tilde{\delta h}$, which satisfies

\begin{equation}\label{zeromode}
L_2[\tilde{\delta h}](\tilde{x}) \equiv
   \tilde{\delta h}''+
   p_t\;\tilde{\delta h}'-
   \tilde{x}\;\tilde{\delta h}-
   3\;\tilde{h}^2\;\tilde{\delta h}=0 \;\;.
\end{equation}
The value of $p_t$ was found in two steps. First, we found solutions to
Eq.~(\ref{UVbis}) for a range of $p$ with the relaxation scheme
$\dot{\tilde{h}}=L_1[\tilde{h}]$. We applied then the relaxation scheme
$\dot{\tilde{\delta h}}=L_2[\tilde{\delta h}]$ to the linear
Eq.~(\ref{zeromode}) with
the initial condition $\tilde{\delta h}(t=0,x)=1$. The field relaxed
to $\tilde{\delta h}(t\rightarrow\infty,x)=0$ for $p<p_t$,
and it blew up without limit for $p>p_t$. For $p \approx p_t$,
we observed a long lived localized zero mode structure. The threshold
estimated in this way is
$ p_t = 6.5 \ldots 6.6  \;\;.$

  Defects can be produced for $v>v_t$, where

\begin{equation}
v_t=\left(1+\frac{2^{1/2}\eta^{3/2}\tau_Q^{1/2}}{p_t^{3/2}}\right)^{-1/2}
    \stackrel{\eta\rightarrow\infty}{\approx}
    \frac{p_t^{3/4}}{\eta}
    (\frac{ \eta }{ \tau_Q })^{1/4} \;\;.
\end{equation}

The instability appears because
the eigenvalue of the lowest mode of a linearized fluctuation
operator around $h_v$ passes through zero when $v=v_t$. The passage is
smooth, so we do not expect any discontinuity in the density of defects
as a function of $v$. For the same reason, we expect the threshold
at $v_t$ to be gradually softened with increasing noise temperature
$\theta$. For $v\gg v_t$, the inhomogeneity of the quench is irrelevant,
and the density of defects can be estimated by scaling argument
\cite{zurek1,zurek2} as for a homogeneous quench with a timescale 
$\tau_Q$.

\begin{figure}[h]
\leavevmode
\epsfxsize=3.2truein\epsfbox{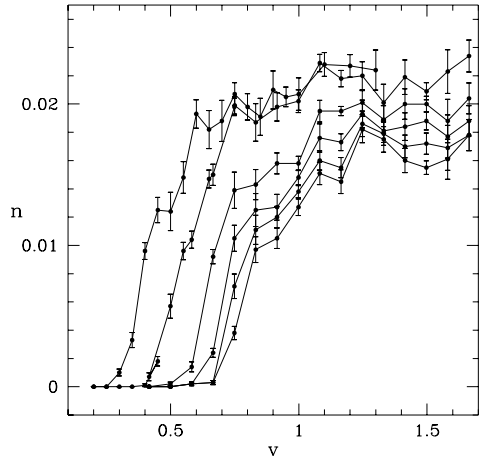}
\caption[figure2]{\label{fig2}
Density of kinks $n$ as a function of velocity $v$
for the linear inhomogeneous quench (\ref{model2}) with $\tau_Q=64$
and $\eta=1$. The predicted threshold is $v_t=0.77$. This cut-off
is achieved for low $\theta$. The plots from top to bottom
correspond to 
$\theta=10^{-1},\, 10^{-2},\, 10^{-4},\, 10^{-6},\, 10^{-8},\,10^{-10}$.}
\end{figure}

This analysis is confirmed by the numerical study of linear quenches shown
in Fig.~\ref{fig2}. For the lowest temperatures, there are no kinks formed below the 
threshold, which for our $\eta=1$ is $v_t \approx 0.77$. However, as temperature
increases from $\theta = 10^{-10}$ to $\theta = 0.1$, kinks begin to appear
at velocities as low as $\sim$0.42. This decrease of the threshold for kink
formation is now more gradual than for the shock wave case of Fig.~\ref{fig1}.

\paragraph*{\bf  CONCLUDING REMARKS.  }

   We found that for both the shock and the linear inhomogeneous quench,
the density of defects is drastically suppressed for quench velocities
lower than the characteristic velocity $v_t\sim\hat{v}$. This prediction 
was verified by numerical simulations for kinks in one dimension.
The theory can be generalized to higher dimensions and to complex
order parameter in a straightforward manner.

   More quantitative understanding of the dependence of the number of 
defects on the quench velocity requires further investigation.
The non-hermitian nature of the operator $L_2[..]$ in Eq.(\ref{zeromode})
should be carefully taken into account. 

   Our prediction that no defects are produced below certain threshold is in
contrast with Ref.\cite{kv}, where some defects are predicted even at very
low velocities (but see Ref.\cite{ekv} for a different conclusion). Formation
of planar solitons in $^3He-A$ \cite{planar}, which takes place at velocities
$10^2 - 10^3$ times lower than the threshold velocity, seems to confirm the
latter theory. Let us however make two remarks:

 1) No uniform temperature gradient was deliberately applied in this
experiment although some non-uniformities could result in local temperature
gradients. There may have existed such regions of the sample where the
gradient was null or negligible. Solitons could be easily created in these
areas. This explanation can be verified in the same experimental set up;
application of a sufficiently strong uniform temperature gradient across
the sample should suppress any soliton formation.

 2) Moreover, in three dimensions, the situation is slightly more complicated.
Straightforward generalization of our analysis would rule out formation of
zero-dimensional defects, as well as one and especially two dimensional 
defects parallel or askew to the direction of the quench front. However, such
extended string and domain-wall like structures could still ``grow" in the
direction perpendicular to the front (and parallel to the direction of its
propagation), providing that their seeds exist, say, along the wall where the
symmetry is broken first. Strings would grow from some seeds and
antistrings from the other. The growth of individual (anti-)strings
would not be perfectly perpendicular to the front, they would be
wandering around in a chaotic manner. From time to time the growing ends of
a string and an antistring would meet and coalesce to form a "jumping rope"
with its other ends anchored at the original seeds. For global strings
(like vortices in $^4He$)
the coalescences are accelerated by a long range mutual string-antistring
attraction. Such a rope would shrink to the original wall thus disappearing
from the bulk. At some stage all possible coalescences would have already
taken place leaving only a net number of say, strings, proportional
to the square root of the number of seeds. Some of these survivors
would be forced by their mutual repulsion to terminate on the side walls.
Only a fraction of them would reach the opposite wall spanning
through the bulk where they can be unambiguously detected. Their density is
likely to be, in any case, orders of magnitude below the estimates based on
\cite{zurek1,zurek2}. Similar argument applies to membrane-like solitons
in \cite{planar}.

  We think that inhomogeneity  is a factor which may also need to be taken
into account in interpretation of the recent negative $^4He$ experiment
\cite{Dodd}.

\acknowledgements

We thank M.E. Dodd, P.V.E. McClintock, and G.E. Volovik for discussions.
Work supported in part by NSF PHY 96-01413, 93-57219 (NYI) to P.L.,
KBN grant 2 P03B 008 15 to J.D. and DoE grant W-7405-ENG-36.

%
%

%
%

\end{document}